\pdfoutput=1 
\ifdefined\IsLLNCS
\documentclass{llncs}
\usepackage{amsfonts,amsmath,amssymb,boxedminipage,color,url,nccmath}
\else
\documentclass[11pt]{article}
\usepackage{amsfonts,amsmath,amssymb,boxedminipage,color,url,fullpage,nccmath}
\fi
\usepackage{enumerate,paralist}
\usepackage[hypcap=false]{caption}
\usepackage{lmodern}
\usepackage[bottom]{footmisc} 
\usepackage[pageanchor=false,pdfstartview=FitH,colorlinks,linkcolor=blue,filecolor=blue,citecolor=blue,urlcolor=blue]{hyperref} 
\ifdefined\IsLLNCS\else
\usepackage{amsthm} 
\fi
\usepackage{aliascnt}
\usepackage[numbers]{natbib} 
\usepackage{cleveref}
\usepackage{xspace}
\usepackage{xstring}
\usepackage{float}
\usepackage{color,soul}
\usepackage{tikz}\usetikzlibrary{arrows}
\usetikzlibrary{decorations.markings}
\usepackage{subcaption}
\ifdefined\IsLLNCS
\fi
\newcommand{\remove}[1]{}

\newcommand{\TLLNCS}[2]{\ifdefined\IsLLNCS#1\else #2 \fi}

\ifdefined\IsDraft
    \newcommand{\authnote}[2]{{\bf [{\color{red} #1's Note:} {\color{blue} #2}]}}
\else
    \newcommand{\authnote}[2]{}
\fi

\ifdefined\IsDraft
    
\else
    
\fi

\ifdefined\IsDraft
    \newcommand{\deleted}[2]{{~\textbf{Deleted (#1):}~{\color{red} #2 }}}
\else
    \newcommand{\deleted}[2]{}
\fi
\ifdefined\IsDraft

\else

\fi

\newcommand{\sdotfill}{\textcolor[rgb]{0.8,0.8,0.8}{\dotfill}} 

\newcommand{\Ensuremath}[1]{\ensuremath{#1}\xspace}
\newcommand{\MathAlg}[1]{\mathsf{#1}}

\newcommand{\MathAlgX}[1]{\Ensuremath{\MathAlg{#1}}}

\newcommand \mycaption {\small }     
\newcommand \mylabel {}

\ifdefined\IsLLNCS
    \newenvironment{nfbox}[3]{
    \renewcommand \mycaption {#1}
    \renewcommand \mylabel {#2}
    \begin{center}\small
    \begin{tabular}{|ll|}
    \hline
    \hspace{.3ex}
    \begin{minipage}{.97\linewidth}
         \vspace{0.5ex}
         #3}
         {\smallskip
         \captionof{figure}{\mycaption}
         \label{\mylabel}
     \end{minipage}
     &\hspace{.3ex} \\
     \hline
     \end{tabular}
     \end{center}    
    }
\else
    
\fi

\newcommand{\eqdef}{:=}

\newcommand{\Supp}{\operatorname{Supp}}

\newcommand{\class}[1]{\mathrm{#1}}

\renewcommand{\P}{\class{P}}

\renewcommand{\cref}{\Cref}

\TLLNCS{
\newaliascnt{claiml}{theorem}
\newtheorem{claiml}[claiml]{Claim}
\aliascntresetthe{claiml}

\renewenvironment{claim}{\begin{claiml}}{\end{claiml}}
}
{
\newtheorem{theorem}{Theorem}[section]

\newaliascnt{lemma}{theorem}
\newtheorem{lemma}[lemma]{Lemma}
\aliascntresetthe{lemma}

\newaliascnt{claim}{theorem}
\newtheorem{claim}[claim]{Claim}
\aliascntresetthe{claim}

\newaliascnt{corollary}{theorem}
\newtheorem{corollary}[corollary]{Corollary}
\aliascntresetthe{corollary}

\newaliascnt{proposition}{theorem}

\aliascntresetthe{proposition}

\newaliascnt{conjecture}{theorem}

\aliascntresetthe{conjecture}

\newaliascnt{definition}{theorem}
\newtheorem{definition}[definition]{Definition}
\aliascntresetthe{definition}

\newaliascnt{remark}{theorem}
\newtheorem{remark}[remark]{Remark}
\aliascntresetthe{remark}

\newaliascnt{example}{theorem}

\aliascntresetthe{example}
}

\crefname{lemma}{Lemma}{Lemmas}
\crefname{figure}{Figure}{Figures}
\crefname{claim}{Claim}{Claims}
\crefname{corollary}{Corollary}{Corollaries}
\crefname{proposition}{Proposition}{Propositions}
\crefname{conjecture}{Conjecture}{Conjectures}
\crefname{definition}{Definition}{Definitions}
\crefname{remark}{Remark}{Remarks}
\crefname{exmaple}{Example}{Examples}

\newaliascnt{construction}{theorem}

\aliascntresetthe{construction}
\crefname{construction}{Construction}{Constructions}

\newaliascnt{fact}{theorem}

\aliascntresetthe{fact}
\crefname{fact}{Fact}{Facts}

\newaliascnt{notation}{theorem}

\aliascntresetthe{notation}
\crefname{notation}{Notation}{Notation}

\crefname{equation}{Equation}{Equations}

\newaliascnt{proto}{theorem}

\newtheorem{proto}[proto]{Protocol}

\aliascntresetthe{proto}
\crefname{proto}{protocol}{protocols}

\newaliascnt{algo}{theorem}

\aliascntresetthe{algo}
\crefname{algo}{algorithm}{algorithms}

\newaliascnt{expr}{theorem}
\newtheorem{expr}[expr]{Experiment}
\aliascntresetthe{expr}
\crefname{experiment}{experiment}{experiments}

\def\FullBox{$\Box$}
\def\qed{\ifmmode\qquad\FullBox\else{\unskip\nobreak\hfil
\penalty50\hskip1em\null\nobreak\hfil\FullBox
\parfillskip=0pt\finalhyphendemerits=0\endgraf}\fi}

\def\qedsketch{\ifmmode\Box\else{\unskip\nobreak\hfil
\penalty50\hskip1em\null\nobreak\hfil$\Box$
\parfillskip=0pt\finalhyphendemerits=0\endgraf}\fi}

\renewcommand{\Pr}{{\mathrm {Pr}}}

\newcommand{\Ac}{\MathAlgX{A}}

\newcommand{\Bc}{\mathsf{B}}
\newcommand{\Cc}{\mathsf{C}}

\newcommand{\party}[1]{%
    \IfEqCase{#1}{%
        {1}{\Ac}
        {2}{\Bc}
        {3}{\Cc}
    }[\PackageError{\party}{Undefined option to party: #1}{}]%
}%

\mathchardef\mhyphen="2D

\renewcommand{\P}{\class{P}}

\renewcommand{\cref}{\Cref}

\crefname{lemma}{Lemma}{Lemmas}

\def\FullBox{$\Box$}
\def\qed{\ifmmode\qquad\FullBox\else{\unskip\nobreak\hfil
		\penalty50\hskip1em\null\nobreak\hfil\FullBox
		\parfillskip=0pt\finalhyphendemerits=0\endgraf}\fi}

\def\qedsketch{\ifmmode\Box\else{\unskip\nobreak\hfil
		\penalty50\hskip1em\null\nobreak\hfil$\Box$
		\parfillskip=0pt\finalhyphendemerits=0\endgraf}\fi}

\title{Super-Spreaders Out, Super-Spreading In:
		The Effects of Infectiousness Heterogeneity and Lockdowns on Herd Immunity}

   \author{
	   	Jhonatan Tavori
	   	\thanks{Blavatnik School of Computer Science, Tel Aviv University, Israel. E-mail: \texttt{jhonatan.tavori@cs.tau.ac.il}}
   		\and 
   		Hanoch Levy
   		\thanks{Blavatnik School of Computer Science, Tel Aviv University, Israel. E-mail: \texttt{hanoch@tauex.tau.ac.il}}}

\begin{document}

\sloppy
\maketitle

\begin{abstract}
	Recently,  \cite{oz2020heterogeneity} has proposed that heterogeneity of infectiousness (and susceptibility) across individuals in infectious diseases, plays a major role in affecting  the Herd Immunity Threshold (HIT).  Such heterogeneity has been observed in COVID-19 and is recognized as \textit{overdispersion} (or ”super-spreading”). 
	The model of \cite{oz2020heterogeneity} suggests that super-spreaders contribute significantly to the effective reproduction factor, $R$, and that they are likely to get infected and immune early in the process. Consequently, under
	$R_0 \approx 3$ (attributed to COVID-19), the Herd Immunity Threshold (HIT) is as low as 5\%, in contrast to 67\% according to the traditional models \cite{1926531, anderson1992infectious, fine2011herd, randolph2020herd}.  
	
	This work follows up on \cite{oz2020heterogeneity} and proposes that heterogeneity of infectiousness (susceptibility) has two “faces” whose mix affects dramatically the HIT: 
	(1) \textit{Personal-Trait-}, and 
	(2) \textit{Event-Based-} Infectiousness (Susceptibility). 
	The former is a personal trait of specific individuals (\textit{super-spreaders}) and is nullified once those individuals are immune (as in \cite{oz2020heterogeneity}). 
	The latter is event-based (e.g cultural \textit{super-spreading} events) and remains effective throughout the process, even after the super-spreaders immune. 
	We extend \cite{oz2020heterogeneity}’s model to account for these two factors, analyze it and
	conclude that the HIT is very sensitive to the mix between (1) and (2), and under $R_0 \approx 3$ it can vary between 5\% and 67\%. 
	Preliminary data from COVID-19 suggests that herd immunity is not reached at 5\%. 
	
	We address operational aspects and analyze the effects of lockdown strategies on the spread of a disease. 
	We find that herd immunity (and HIT) is very sensitive to the lockdown type. 
	While some lockdowns affect positively the disease blocking and increase herd immunity, others have \textit{adverse} effects and \textit{reduce} the herd immunity.
\end{abstract}


\section{Introduction}
In 1923 Topley and Wilson described experimental epidemics in which the rising prevalence of immune individuals would end an epidemic. They named this phenomenon as \textit{"Herd-Immunity"} \cite{topley1923spread}. Once the Herd Immunity Threshold (HIT, measured in fractions of the population that got immune) is surpassed, then the effective reproduction number, $R$, reduces below $1$ and the number of infection cases decreases. The exact value of this threshold is an important measure used in infectious disease control and immunization, and its estimation for the COVID-19 disease are used by governments worldwide in determining policies to fight against the current pandemic. 

A recent study of Oz, Rubinstein and Safra \cite{oz2020heterogeneity} proposed a new model for the spreading of infectious diseases, such as COVID-19. 
Under that model and the assumption that the basic reproduction number, $R_0$, of COVID-19 is approximately $3$, the HIT is approximately 5\%, namely, when 5\% of the population is infected herd immunity is reached.
This estimation was in contrast to the allegedly “axiomatic” cutpoint of $HIT \approx 67\%$ assumed for COVID-19 \cite{randolph2020herd}.

Preliminary data from COVID-19 suggests that herd immunity is not reached at the approximate  5\% fraction expected by \cite{oz2020heterogeneity}. 
For Example, the US states of North Dakota, South Dakota, Iowa, Utah and Tennessee with at least 12\%, 11.5\%, 9.5\%, 9.5\% and 9\% infected, respectively. Four of them have current $R > 1$ \cite{currenrRt, Worldometers}.

The model of \cite{oz2020heterogeneity} is based on the observation \cite{pastor2015epidemic,rock2014dynamics, tkachenko2020persistent} that the epidemic spread network is not homogeneous, where distinct individuals are infectious (likely to infect others) and susceptible (likely to become infected themselves) in various degrees.
The heterogeneity of these values among individuals is recognized as 
\textit{overdispersion} or \textit{super-spreading} (\textit{super-spreaders} are a class of individuals whose secondary infection rate is very high \cite{smith2005}). 
The estimates for the COVID-19 pandemic fits this property and asserts that between 5\% to 10\% of the infected individuals cause 80\% of the secondary infections \cite{ endo2020estimating, miller2020full}.

Furthermore, a correlation between the infectiousness and susceptibility of each individual have a drastic effect on the over-time reduction of the effective reproduction number, $R$, under the spreading model of \cite{oz2020heterogeneity}. 
The heterogeneity and correlation of these parameters yields that the "super-spreaders" are extremely likely to get infected and develop immunity in an early stage of the pandemic process.
\cite{oz2020heterogeneity}'s 5\% estimation of the percentage of the population that contract the disease before herd immunity is reached was based on these properties.

In this work, we follow-up on \cite{oz2020heterogeneity} regarding the effect of infectiousness heterogeneity on the HIT. However, we propose that infectiousness (and susceptibility) should be classified into two inherently different types: (1) \textit{Personal-Trait Infectiousness (Susceptibility)}, and (2)  \textit{Event-Based Infectiousness (Susceptibility)}. We will refer to the combination of the infectiousness and the susceptibility as \textit{Spreading}.

The first type stems from traits of an individual. The second type relates to social events in which every individual may participate, regardless of its personal traits.   
To demonstrate these two types, consider, for example, a package-delivery person and compare it with an academic researcher.  
During a single day, the delivery person has interactions with tens or hundreds of people, and therefore has high personal-trait infectiousness. On the other hand, the researcher may work most of the time in his/her office or interact with a small research group and as a result has a lower personal-trait infectiousness.
Yet, both of them may participate in a social-gathering event (such as a concert, a wedding, or "just" a family birthday party). During such an event, both have approximately the same amount of interaction (which may be quite large), and therefore have the same event-based spreading degree, regardless of their personal traits.
It is important to note that personal-trait infectiousness relate not only to the social behaviour of the individual; it may also relate to his/hers biological properties  (e.g., his/her body reproduces a virus faster and therefore he/she is more infectious).

We assume that the likelihood parameters of each individual consist each of the sum of two parameters.
(1) $S_p(a)$ and $I_p(a)$ which are the personal-trait susceptibility and infectiousness parameters of $a$, respectively. Those values reflect personal traits and are drawn once (pandemic beginning) and remain constant throughout, exactly as in \cite{oz2020heterogeneity}. 
(2) $S_e$ and $I_e$ which are the event-based (cultural) susceptibility and infectiousness parameters. Those values reflect occasional event-based spreading which is probabilistically redrawn for each individual at every step of the pandemic.
We assume that the likelihood of $a$ to be infected is:
$$ S(a) = p \cdot S_p(a) + (1 - p) \cdot S_e$$
and the likelihood of $a$ to infect others is
$$ I(a) = p \cdot I_p(a) + (1 - p) \cdot I_e,$$
where $p$ determines the weight of each infection type.
The symmetry between S(a) and I(a) is similar to that of \cite{oz2020heterogeneity} and stems from the assumption that infectiousness level and susceptibility level are proportional to the level of interaction the individual has with others or to its biological properties.   

We use this model to analyze the progression of an infectious disease and the value of $R(n)$, the effective reproduction number, as a function of the fraction of population infected. 
We show that 
the contribution of the personal-trait spreading drops sharply at early stages of the disease, as the super-spreaders are likely to contract the disease and develop immune at early stages of the process.
On the other hand, the contribution of the event-based spreading drops much more slowly, and is affected very little at early stages, as its reduction is proportional to the decrease of the susceptible population size.
In other words -- when super-spreaders are out (event-based) super-spreading is in.
Hence, $R(n)$ may remain at high values even after the super-spreaders population is fully immune, as opposed to \cite{oz2020heterogeneity}. This results in a slower decay of $R(n)$ and leads to a higher value of the Herd Immunity Threshold.

We show that for COVID-19 the Herd Immunity Threshold depends on the \textit{mix} ($p$) between the weights of the personal-trait and event-based spreading. 
In particular, we prove that the $HIT \approx 5\%$ estimate of \cite{oz2020heterogeneity} holds when assuming only personal-trait spreading, and that the traditional prediction of $HIT \approx 67\%$ holds when assuming only event-based spreading. 

Having established a formula expressing $R(n)$, we address operational aspects and analyze the effects of lockdowns on the Herd Immunity Threshold.  
Lockdowns, of a variety of variants, have been enforced worldwide in order to fight the COVID-19 pandemic. While lockdowns might have immediate impact such as collapse of the effective reproduction and suppression of infections and mortality, they have long-term impact as well.
We discuss two different lockdown policies: (1) An Event-based spreading targeted lockdown (e.g shut down of cultural events). (2) A Personal-trait spreading targeted lockdown (e.g. restricting daily/professional activities).
We analyze the effect of these policies on the composition of the infected population, on the effective reproduction number, and on the Herd Immunity Threshold (HIT). 
We show that while a lockdown (or a sequence of lockdowns) which is targeted at event-based spreading reduces the disease spread by decreasing the HIT, a lockdown (or a sequence of lockdowns) which is targeted at personal-trait spreading will act adversely and will \textit{increase} the disease spread by increasing the HIT.  

The rest of the paper in organized as follows: 
In Section 2 we formally describe our model, and present the effective reproduction number which plays a major role in the analysis.
In Section 3 we develop an expression for $R(n)$, and calculate the Herd Immunity Threshold for a general-case disease. We examine the result in a numerical discussion based on COVID-19 spreading distributions.
Then, in Section 4, we study various lockdown policies, and analyze their effect on the Herd Immunity Threshold.
Finally, concluding remarks are given in Section 5.

\section{The Disease Spread Model}

In this section, we present our model for the spreading of infectious disease accounting for  heterogeneity of infectiousness/susceptibility. 
We extend the model of \cite{oz2020heterogeneity} and propose that there exists an event-based infection factor in addition to the factors described in their model.
Our analysis begins with a certain number of infected individuals. 
We measure the spread of the disease as a function of the number of individuals who got infected.
Specifically, we index the individuals by the order they are infected and have $R(n)$ denote the effective reproduction number associated with the $n$th infected individual. 
Namely, the event whereby the $n$th individual gets infected is the $n$th event (or step $n$). 
We use $n$ also to denote the \textit{step of the disease}.

Measuring the spread of the disease as a function of the infected population size will be useful in deriving the Herd Immunity Threshold (HIT) of the disease, namely the fraction of the population that gets infected prior to reaching $R(n) \le 1$.

\subsection{Susceptibility and Infectiousness}
We follow the model of \cite{oz2020heterogeneity} and assign to each individual $a$ a personal-trait-susceptibility parameter $S_p(a)$ and a personal-trait-infectiousness parameter $I_p(a)$ drawn from some probability distributions. 
Those values quantify how likely $a$ is to be infected and infect others, respectively, according to its \textit{personal} traits.
The values of $S_p(a)$ and $I_p(a)$ accompany $a$ throughout the entire progress of the disease, and remain at the same values.
We follow \cite{oz2020heterogeneity} and define the average conditional infectiousness $\varphi(s)$. In our case, it is logical to parametrize  $\varphi(s)$ only by the personal-trait susceptibility and infectiousness:
\begin{equation}
\varphi(s) := \mathbb{E}_{S_p(a) = s} \left[  I_p(a) \right].
\end{equation}

As was discussed, the heterogeneity of the spreading values of the population will play a major role in our analysis. Hence, we will measure: \footnote{For continuous $s$ , Eq. (\ref{defrho}) should be considered as a density function. }
\begin{equation}\label{defrho}
\rho(s, n) := \Pr \left[ S_p(a) = s \Big\vert a \in H_{n} \right] 
\end{equation}
where $H_{n}$ is the healthy population at step $n$.

In addition, and beyond the model of \cite{oz2020heterogeneity}, we assign an event-based infectiousness parameter and event-based susceptibility parameter to each individual. 
Those values are subject to change through the progress of the disease. 
At step $i$ we assign to $a$ $S_e^i(a)$, the event-based-susceptibility parameter and $I_e^i(a)$,  the event-based-infectiousness parameter, both are random variables whose realizations hold only for iteration $i$. 
Since the values of $S_e, I_e$ measure the event-based-spreading of the society (assigned to its individuals at a given time), they are drawn from probability distributions that are common for the entire population, denote them by $\Lambda_S$ and $\Lambda_I$. 

The \textit{susceptibility} of $a$ at step $i$, which is the likelihood of $a$ to be infected, is:
\begin{equation}\label{s_def}
S^i(a) = p \cdot S_p(a) + q \cdot S_e^i(a),
\end{equation}
and the \textit{infectiousness} of $a$, which is the likelihood of $a$ to infect others is
\begin{equation}\label{i_def}
I^i(a) = p \cdot I_p(a) + q \cdot I_e^i(a)
\end{equation}
where $0 < p < 1$ and $q = 1-p$. The value of $p$ determines the \textit{mix} between the spreading types. We call $p$ (and respectively, $q$) the weights of the personal-trait (and respectively, event-based) spreading of the disease. As will be seen later, the value of $p$ will have a drastic effect on the Herd Immunity Threshold.
Note that the special case where $p=1$ gives exactly the model of \cite{oz2020heterogeneity}. 

Under this model, the probability that $a$ will be infected at step $n$, assuming that $a$ was healthy at step $n-1$ is: 
\begin{equation}\label{prob_h}
\Pr \left[ \: a \text{ is the $n$th infected } \Big\vert \: a \text{ is health in step $n-1$} \: \right] =
\frac{S^{n-1}(a)}{\sum_{b \in H_{n-1}}S^{n-1}(b)}
\end{equation}

\subsection{Basic and Effective Reproduction Number}
The \textit{basic reproduction number}, $R_0$, is a measure of how transferable a disease is. It is defined as the expected number of secondary cases produced by a single (typical) infection in a completely susceptible population (whose size is $N_0$).

In reality, varying proportions of the population are immune to any given disease at any given time. Hence, as in \cite{oz2020heterogeneity}, we will measure the \textit{effective reproduction number}, $R(n)$, which is defined as the expected number of infections directly generated by the $n$th infected individual.  

\begin{equation}\label{defrstraight}
R(n) = \mathbb{E} \left[  \verb|#|  \text{ of individuals that will be infected by the $n$th infected}\right]
\end{equation}
where the expectation is taken over the $n$th individual to be infected. 
As in \cite{oz2020heterogeneity}, Eq. (\ref{defrstraight}) equals to\footnote{The expectation is taken over all possible scenarios of infection.}
$$R(n)=\mathbb{E}[I^{n}(a)\cdot\sum_{b\neq a}S^{n}(b)]=$$
$$=\sum_{a \in H_{n-1}} \frac {S^{n-1}(a)}{\sum_{b \in H_{n-1}}S^{n-1}(b)}\cdot I^{n}(a)\cdot\sum_{ b \neq a \in H_{n-1}}S^{n}(b)=$$
$$=\sum_{a \in H_{n-1}} \frac{\sum_{b \neq a \in H_{n-1}}S^{n}(b)}{\sum_{b \in H_{n-1}}S^{n-1}(b)}\cdot S^{n-1}(a)\cdot I^{n}(a)$$
this can be approximated by: 
\begin{equation}\label{R_approx}
R(n) \approx \sum_{a\in H_{n-1}}S^{n-1}(a)\cdot I^{n}(a).
\end{equation}
Using Eq. (\ref{s_def}), (\ref{i_def}) and (\ref{R_approx}) we have:
\begin{equation}\label{R_expers}
R(n) \sim N(n) \cdot \int  \rho(\sigma, n) \cdot \left(p\cdot \sigma + q\cdot \lambda_{S} \right) \cdot \left(p\cdot \varphi(\sigma)+q\cdot \lambda_{I} \right) d\sigma
\end{equation}
where $N(n)$ is the size of the healthy population at step $n$ and $\lambda_{I}, \lambda_{S}$ are the means of $\Lambda_{I}, \Lambda_{S}$, respectively.

\section{Reaching Herd Immunity}
In this section, we analyze the changes in the  composition of the population through the spread of the disease, and the decrease of $R(n)$ as the fraction of the population that contracted with the disease increases. 
We prove the following theorem:

\begin{theorem}[General Case Herd Immunity Threshold]\label{thm1}
	For any $\delta$ when
	\begin{equation}\label{thm_precent}
	1 - \int\rho(\sigma)\cdot\exp\left(-\delta\cdot(p\cdot\sigma+q\cdot\lambda_{S})\right)d\sigma
	\end{equation}
	fraction of the population is infected, the effective reproduction number, $R()$, will be reduced by a factor of
	\begin{equation}\label{thm_r}
	\frac{\int\rho(\sigma,0)\cdot\exp\left(-\delta \cdot \left(p\cdot\sigma+q\cdot\lambda_{S}\right)\right)\cdot \left(p\cdot \sigma + q\cdot \lambda_{S} \right) \cdot \left(p\cdot \varphi(\sigma)+q\cdot \lambda_{I} \right)d\sigma}{\int\rho(\sigma,0) \cdot \left(p\cdot \sigma + q\cdot \lambda_{S} \right) \cdot \left(p\cdot \varphi(\sigma)+q\cdot \lambda_{I} \right)d\sigma}
	\end{equation}
	relatively to the basic reproduction number, $R_0$. The threshold for herd immunity is when the value of the effective reproduction number is $1$.
\end{theorem}
Having this expression for the change in the effective reproduction number for a general distribution, we will later use the special case of the Gamma distribution with estimated parameters for COVID-19 \cite{endo2020estimating, smith2005, lourenco2020impact} and inspect the HIT values for different $p$ values.

\subsection{Proof of Theorem \ref{thm1} for A General Spreading Distribution}
\label{subsec31}

In order to prove Theorem \ref{thm1}, we establish the following claim:
\begin{claim}[The likelihood of an individual to be infected]\label{lem32}
	For a person $a$,
	\begin{equation}\label{a_healty}
	\Pr[a \text{ is healthy at round }  n] \approx \exp\left(-\beta(n)\cdot(p\cdot S_{p}(a)+q\cdot\lambda_{S})\right)
	\end{equation}
	where 
	\begin{equation}\label{betadef}
	\beta(n) = \sum_{{i=0}}^{{n-1}} \frac{1}{N(i) \cdot \mathbb{E}_{b \sim H_{i}}[S^i(b)]}.
	\end{equation}
\end{claim}

\begin{proof}[Proof of Claim \ref{lem32}]
	The proof follows the proof of Claim I provided in \cite{oz2020heterogeneity} with modifications required for our extended model. The proof is based on using Eq. (\ref{prob_h}) and obtaining: 
	$$\Pr[a \text{ is healthy at step } n] = $$
	\begin{equation}\label{eqclmpra}
	\left(1-\frac{S^{n-1}(a)}{{N(n-1) \cdot \mathbb{E}_{b \sim H_{n-1}}[S^{n-1}(b)]}}\right)
	\cdot 
	\Pr[a \text{ is healthy at step } n-1].
	\end{equation}
	The rest of the proof consists of algebraic manipulations of (\ref{eqclmpra}) and the full details are given in Appendix \ref{appendA2}. 
\end{proof}

We next establish two supporting lemmas, corresponding to equations (3.6) and (3.7) in \cite{oz2020heterogeneity}, and conclude with the proof of Theorem \ref{thm1}.

\begin{lemma}[Heterogeneity of the population during the process]\label{lem33}
	For any $s \in \Supp(S_p)$,
	$$\rho(s,n) \approx \frac{\rho(s,0)\cdot\exp\left(-\beta(n)\cdot p\cdot s\right)}{\int\rho(\sigma,0)\cdot\exp\left(-\beta(n)\cdot p\cdot\sigma\right)d\sigma}$$
\end{lemma}

\begin{proof}[Proof of Lemma \ref{lem33}]
	By definition 
	\begin{equation}
	\rho(s,n)=\rho(s,0)\cdot\frac{\Pr[a\text{ is healthy at round \ensuremath{n}\ensuremath{\vert S_{1}(a)=s}]}}{\int\rho(\sigma,0)\cdot\Pr[a\text{ is healthy at round \ensuremath{n}\ensuremath{\vert S_{1}(a)=\sigma}]}d\sigma}.
	\end{equation}
	Using Eq. (\ref{a_healty}),
	$$ \rho(s,n) \approx\frac{\rho(s,0)\cdot\exp\left(-\beta(n)\cdot(p\cdot s+q\cdot\lambda_{S})\right)}{\int\rho(\sigma,0)\cdot\exp\left(-\beta(n)\cdot(p\cdot\sigma+q\cdot\lambda_{S})\right)d\sigma}=\frac{\rho(s,0)\cdot\exp\left(-\beta(n)\cdot p\cdot s\right)}{\int\rho(\sigma,0)\cdot\exp\left(-\beta(n)\cdot p\cdot\sigma\right)d\sigma}.$$
\end{proof}

\begin{lemma}[The size of the susceptible population]\label{lem34}
	For any $n \in [N_0]$,
	\begin{equation}\label{n_size}
	N(n)\approx N_0\cdot\int\rho(\sigma,0)\cdot\exp\left(-\beta(n)\cdot(p\cdot\sigma+q\cdot\lambda_{S})\right)d\sigma. 
	\end{equation}
	where $N(n)$ is the size of the susceptible population at step $n$, and $N_0$ is the total size of the population.
\end{lemma}

\begin{proof}[Proof of Lemma \ref{lem34}]
	By definition, 
	\begin{equation}
	N(n)\approx N_0\cdot\int\rho(\sigma,0)\cdot\Pr[a\text{ is healthy at round \ensuremath{n}\ensuremath{\vert S_p(a)=\sigma}]} d\sigma. 
	\end{equation}
	Using Eq. \ref{a_healty}, we have Eq. (\ref{n_size}).
\end{proof}

\begin{proof}[Proof of Theorem \ref{thm1}]
	Denote 
	$$r(\sigma) \eqdef \left(p\cdot \sigma + q\cdot \lambda_{S} \right) \cdot \left(p\cdot \varphi(\sigma)+q\cdot \lambda_{I} \right).$$
	Using Eq. (\ref{R_expers}) we know that:
	$$R_0= N_0 \cdot \int  \rho(\sigma, 0) \cdot r(\sigma) d\sigma.$$
	We develop the ratio:
	$$\frac{R(n)}{R_0}=\frac{N(n) \cdot \int  \rho(\sigma, n) \cdot r(\sigma)  d\sigma}{N_0 \cdot \int  \rho(\sigma, 0) \cdot r(\sigma)  d\sigma}.$$
	According to Lemma \ref{lem33} and Lemma \ref{lem34} we can replace the values of $\rho(s,n)$ and the ration $N(n) / N_0$ and have:
	$$\frac{R(n)}{R_0}=\frac{\int\rho(\sigma,0)\cdot\exp\left(-\beta(n)\left(p\cdot\sigma+q\cdot\lambda_{S}\right)\right)d\sigma \cdot \int \frac{\rho(\sigma,0)\cdot\exp\left(-\beta(n)\cdot p\cdot \sigma \right)}{\int\rho(\sigma',0)\cdot\exp\left(-\beta(n)\cdot p\cdot\sigma'\right)d\sigma'} \cdot r(\sigma)  d\sigma}{\int  \rho(\sigma, 0) \cdot r(\sigma)  d\sigma}=$$
	$$ =\frac{ \int {\rho(\sigma, 0)\cdot\exp\left(-\beta(n)\cdot p\cdot \sigma + q \cdot \lambda_{S} \right)} \cdot r(\sigma)  d\sigma}{\int  \rho(\sigma, 0) \cdot r(\sigma)  d\sigma}.$$
	Since the value of expression (\ref{thm_precent}) is given by: 
	$$ 1 - \frac{N(n)}{N_0}$$
	we have that Eq. (\ref{thm_r}) holds.
\end{proof}

\subsection{COVID-19: Analysis and Discussion }
We move to demonstrate the results of Theorem \ref{thm1} on a Gamma distribution with shape and scale parameters $k$ and $\theta$, respectively.
The Gamma distribution was previously attributed to the infectiousness of COVID-2 \cite{smith2005}.
We substitute the estimates for COVID-19: $R_0 \approx 3$ and $k \approx 0.1$. \cite{endo2020estimating,lourenco2020impact,oz2020heterogeneity}.
We assume that the personal-trait-susceptibility and personal-trait-infectiousness of the population are highly correlated. I.e., we set $\varphi(s) = s$.
This stems from assuming that both personal-trait infectiousness and susceptibility levels are correlated to social interaction levels of the individual or to its biological properties.  

\begin{figure}[H]
	\centering
	\includegraphics[width=1\linewidth]{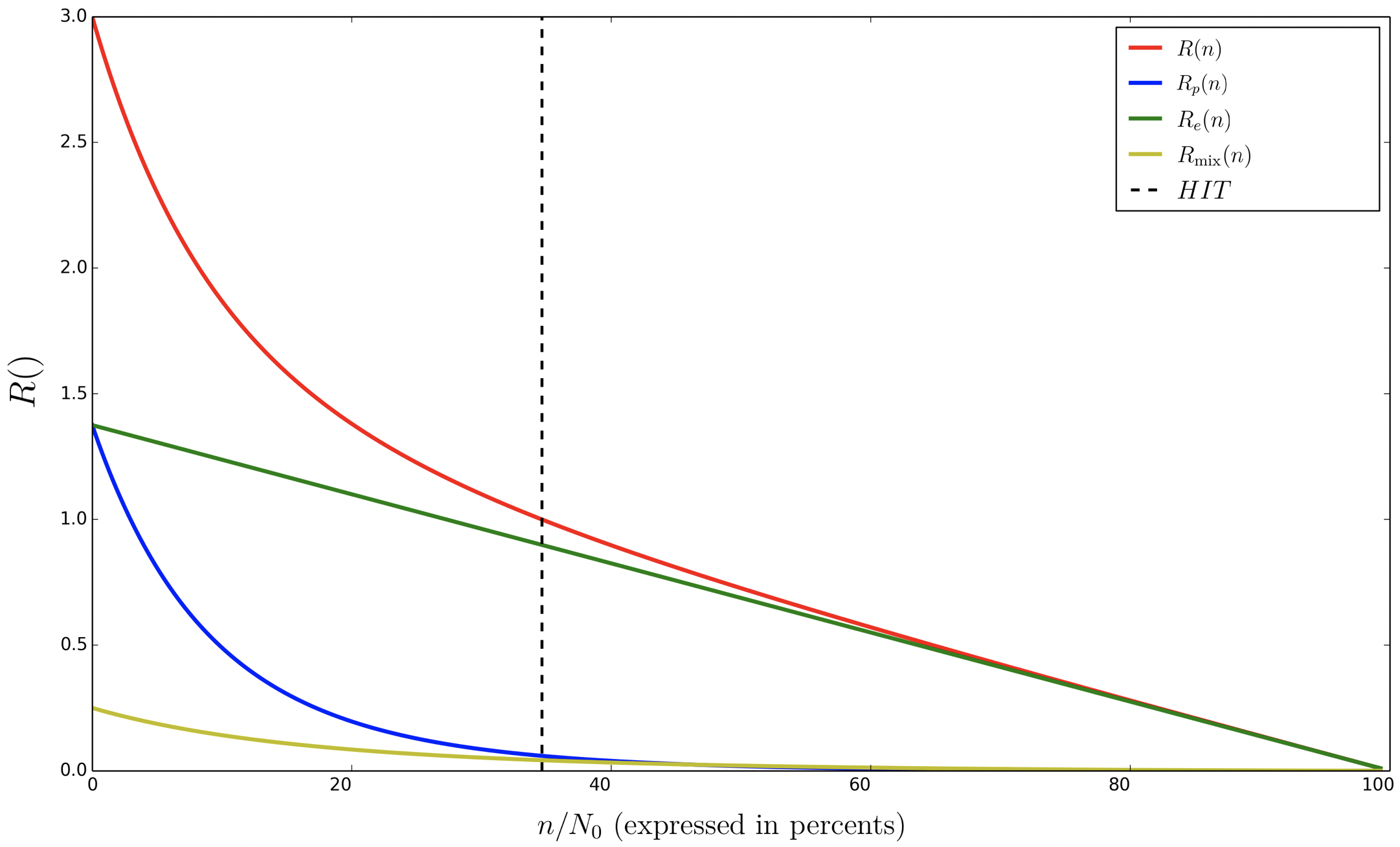}
	\caption{The over-time reduction in the effective reproduction number, $R(n)$, and its contributing factors as a function of $n$, assuming $p=0.5$, $k=0.1$ and $R_0=3$. Note that $n$ (horizontal-axis) is normalized to percentage.
		In red - $R(n)$ (Eq. (\ref{eqnarr1})); In blue - $R_p(n)$  (Eq. (\ref{eqnarr2})); In Green - $R_e(n)$ (Eq. (\ref{eqnarr3}));  In yellow - $R_{mix}(n)$ (Eq. (\ref{eqnarr4}));}
	\label{R_parts}
\end{figure}

In Figure \ref{R_parts} we demonstrate the decay of the effective reproduction number, $R(n)$, and its contributing factors, classified by their spreading types. We plot the following values as a function of the fraction of the infected population. 
\begin{eqnarray}
R(n) & = & N(n) \cdot \int  \rho(\sigma, n) \cdot \left(p\cdot \sigma + q\cdot \lambda_{S} \right) \cdot \left(p\cdot \varphi(\sigma)+q\cdot \lambda_{I} \right) d\sigma \label{eqnarr1} \\
R_p(n) & = & N(n) \cdot \int  \rho(\sigma, n) \cdot p^2 \cdot \sigma \cdot \varphi(\sigma) d\sigma  \label{eqnarr2}  \\ R_e(n) & = & N(n) \cdot \int  \rho(\sigma, n) \cdot q^2 \cdot \lambda_{S} \cdot \lambda_{I} d\sigma  \label{eqnarr3}  \\ R_{mix}(n) & = & R(n) - R_p(n) - R_e(n)  \label{eqnarr4} 
\end{eqnarray}

The blue curve in Figure \ref{R_parts} depicts the contribution of the personal-trait spreading to $R(n)$, while the green curve depicts the contribution of the event-based spreading.
As can be seen, the contribution of the personal-trait spreading drops at early stages of the disease.
This results from the assumption that infectiousness is positively correlated with susceptibility, since if more infectious people are also more susceptible, then they have higher probability to be infected and develop natural immunity much sooner (than the less infectious individuals).
On the other hand, the contribution of the event-based spreading is affected very little at early stages. 
Its reduction is proportional to the decrease of the susceptible population, which is linear in $n$. 
The weight of each spreading type, which is determined by $p$, determines the combined behaviour of $R(n)$ .

\begin{figure}[H]
	\centering
	\includegraphics[width=1\linewidth]{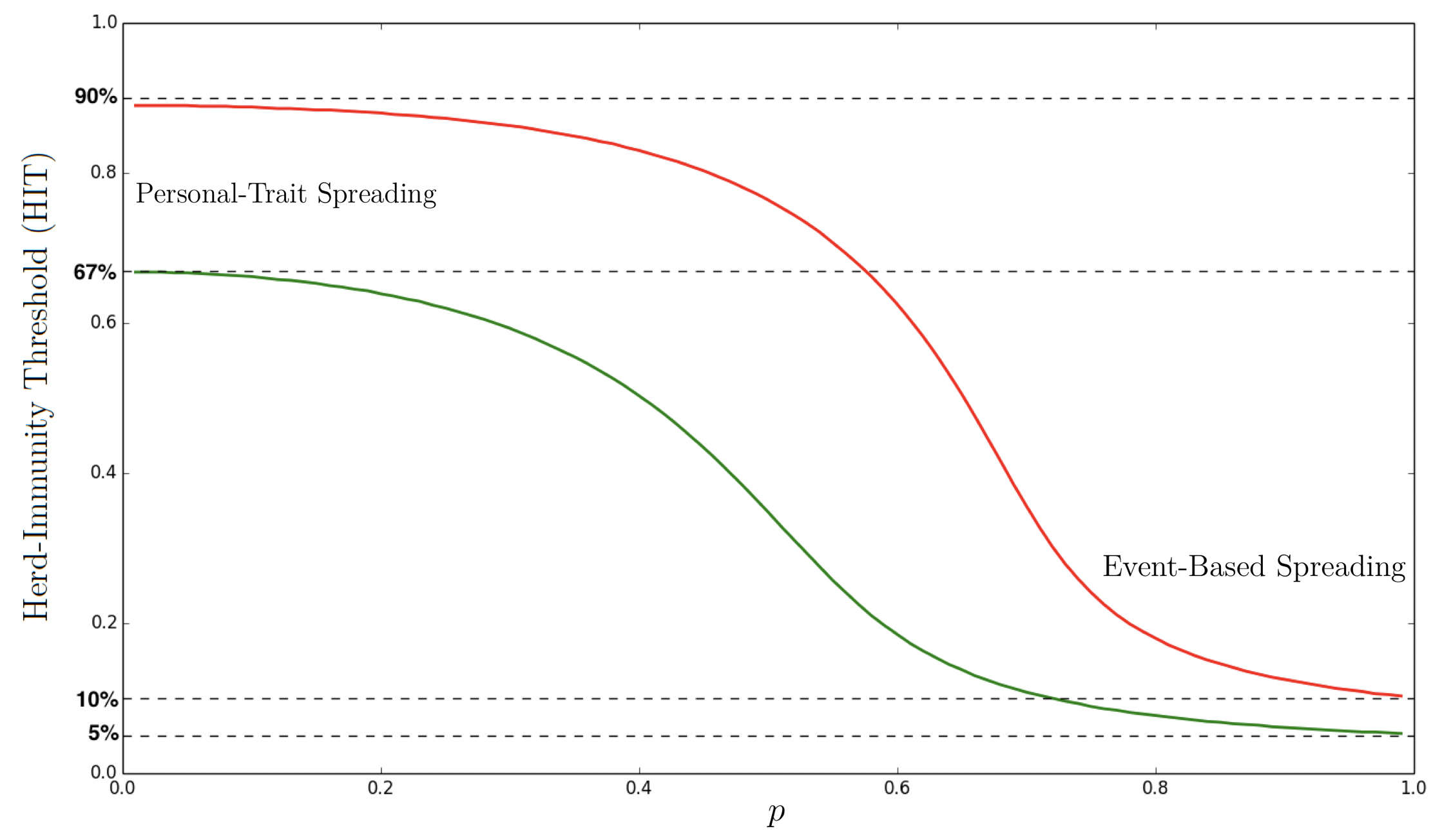}
	\caption{The Herd-Immunity Threshold (HIT) as a function of $p$ assuming Gamma distribution with shape parameter $k=0.1$. In green - $R_0=3$. In red - $R_0 = 9$.}
	\label{herd_by_p_3_9}
\end{figure}

In Figure \ref{herd_by_p_3_9} we plot the Herd Immunity Threshold as a function of $p$. This demonstrates the effect of the weight of the personal-trait spreading on the decay of $R(n)$.

When assuming only event-based-spreading (i.e., $p=0$) the HIT is approximately 
\begin{equation}
\left(1 - \frac{1}{R_0}\right),
\end{equation}
i.e., $67\%$ when $R_0\approx 3$ and $90\%$ when $R_0 \approx 9$. 
This follows the classical models  \cite{1926531, anderson1992infectious, fine2011herd, randolph2020herd}.
On the other hand, assuming only personal-trait-spreading (i.e., $p=1$) the HIT is approximately $5\%$ when $R_0\approx 3$ and $10\%$ when $R_0 \approx 9$. The $5\%$  matches the expected threshold given by \cite{oz2020heterogeneity}. 

\begin{figure}[H]
	\centering
	\includegraphics[width=1\linewidth]{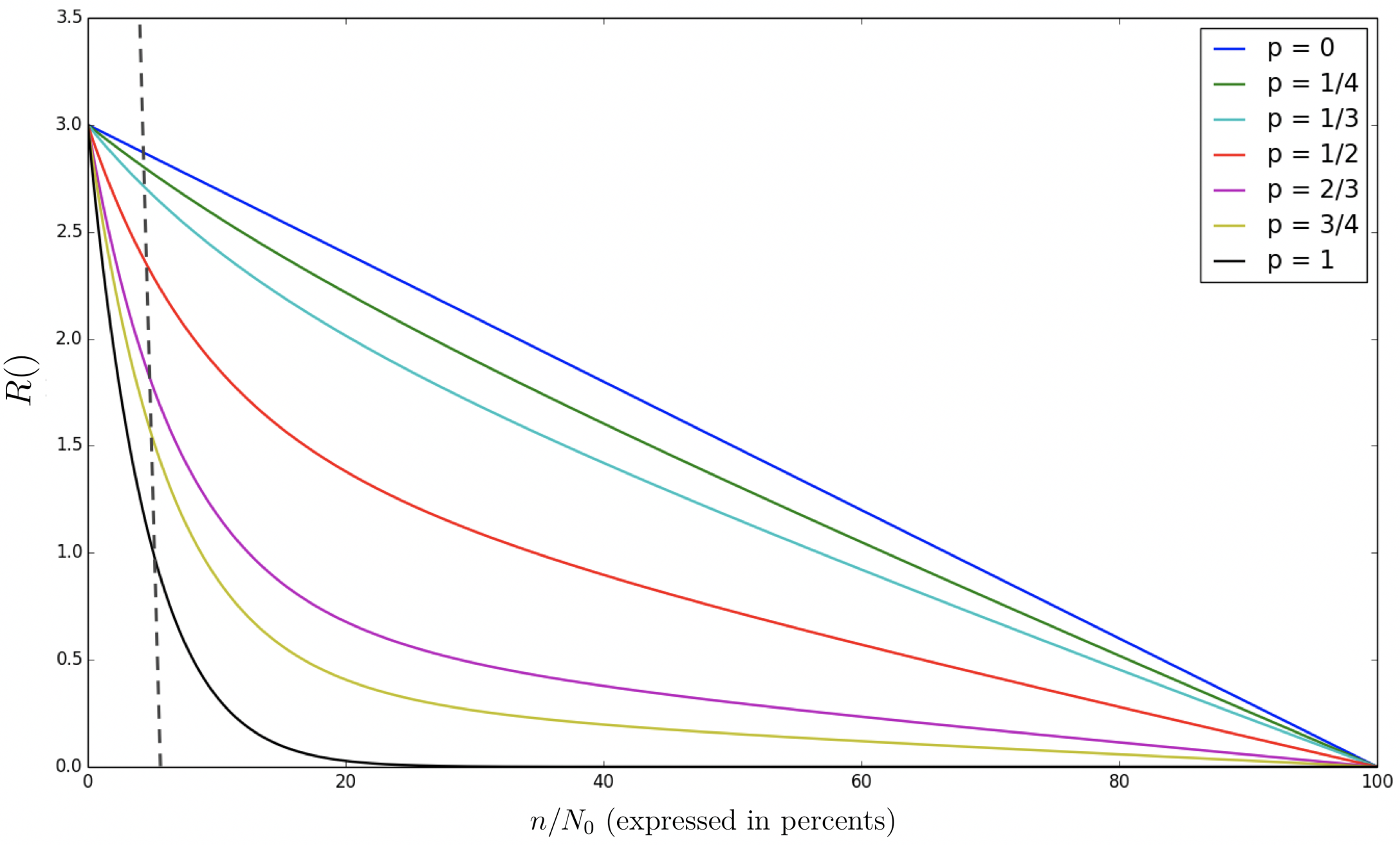}
	\caption{The effective reproduction number $R(n)$ as a function of $n$ throughout the disease spread for different $p$ values where $R_0=3$ and $k=0.1$.  Note that $n$ (horizontal-axis) is normalized to percentage. }
	\label{r_for_ps}
\end{figure}

In reality, the two types of spreading contribute to infections and hence $0 < p < 1$; In order to predict the Herd Immunity Threshold, one has to estimate the value of $p$, given a population. In Figure \ref{r_for_ps} we plot the effective reproduction number, $R(n)$, throughout the spread of the disease for a number of $p$ values. 

Figure \ref{r_for_ps} and the shapes of the various curves can be used to assist in predicting the value of $p$ at \textit{early stages} of a disease. As can be seen, even when a small percentage of the population is infected ($< 5\%$), there is a significant difference between the $R(n)$ values for different $p$'s.
Such prediction can assist in predicting the Herd Immunity Threshold. 

The current data available to us for the current-spreading COVID-19 is biased due to lockdowns of many types, which makes it challenging to estimate $p$ and $q$.
This estimation problem is left open for forthcoming research. 

The effect of lockdown policies on the value of $R(n)$ and the Herd Immunity Threshold is the subject of our next section.

\section{The Effect of Lockdowns}
Lockdown strategies, of a variety of variants, have been used worldwide as a major means to fight an epidemy, specifically COVID-19. The question addressed in this section is how such strategies affect the scope of the disease, namely how they affect the Herd Immunity Threshold.
We show that, depending on its type, a lockdown may either increase or decrease the size of the population infected prior to reaching herd immunity.  
In particular, we show that a personal-trait spreading targeted lockdown acts \textit{adversely} on the efforts to reduce the spread of a disease since it increases the threshold.
In contrast, an event-based spreading targeted lockdown affect positively the disease blocking and reduces the Herd Immunity Threshold.   

\subsection{Lockdown Policies}
Large scale physical distancing measures and moving-around restrictions, often referred to as \textit{lockdowns}, can slow disease transmission by limiting contacts between people.
These days, many restrictions are used worldwide in order to slow down the spread of the COVID-19 pandemic. We classify the restrictions into two inherently different types: 
(1) a Personal-trait spreading targeted lockdown, for example by closing or restricting  workplaces.
(2) an Event-based spreading targeted lockdown, for example by prohibition on cultural events.

Formally, these lockdowns will be defined as follows:
\begin{definition}[Personal-Trait Spreading Targeted Lockdown]
	During a personal-trait spreading targeted lockdown, the infectiousness and susceptibility of an individual $a$ is:
	$$I^i(a) = p_L \cdot I_p(a) + q \cdot I_e^i(a), \:\: S^i(a) = p_L \cdot S_p(a) + q \cdot S_e^i(a),$$
	where $q$ remains the same and $p_L = 0$.
\end{definition}
\begin{definition}[Event-Based Spreading Targeted Lockdown]
	During an event-based spreading targeted lockdown, the infectiousness and susceptibility of an individual $a$ is:
	$$I^i(a) = p \cdot I_p(a) + q_L \cdot I_e^i(a), \:\:  S^i(a) = p \cdot S_p(a) + q_L \cdot S_e^i(a), $$
	where 	$p$ remains the same and $q_L = 0$.
\end{definition}

We will use the following notation. 
Assume  that a lockdown starts at the $n_b$th step of the disease and ends at step $n_e$.
For any $n \in [n_b, n_e]$, let $R_L(n)$ be the expected value of the effective reproduction number during the lockdown;  For any $n > n_e$ let $R_L(n)$ be its expected value after the lockdown is released and $p$ (or $q$) returns to its original value. 
Similarly, let $HIT$ be the Herd Immunity Threshold assuming a "natural" spread of the disease (i.e., no lockdown), and $HIT_L$ be the threshold assuming a lockdown was performed.
Note that the comparison between the natural evolution and the lockdown evolution is based on the number of individuals that contract the disease during those evolutions. I.e., the coupling between the evolutions is done user by user by the order of infection rather than by the time period in which the policies are compared. This coupling method is useful in deriving $HIT$ and $HIT_L$.

At $n_b$ we set $p_L = 0$ or $q_L = 0$. Hence, the value of the effective reproduction number is likely to drop during the lockdown. The question we answer in this section is \textit{if} -- and \textit{what} -- will be the long-term impact of the lockdown, after it is released (at $n>n_e$).
Since at $n_e$ the value of $p$ (or $q$) will increase, $R_L(n)$ is most likely to increase as well. Yet, will it pass $R(n)$? Will it stay below it? Consequently, what will be the impact on the Herd Immunity Threshed?
The next two theorems establish that \textit{any} personal-trait spreading targeted lockdown will increase the HIT while \textit{any} event-based spreading targeted lockdown will decrease the HIT.

\begin{theorem}[HIT of Personal-Trait Spreading Targeted Lockdown]\label{lockpthm}
	Assume that a personal-trait spreading targeted lockdown was performed for $n \in [n_b, n_e]$. Then for any $n > n_e$,
	\begin{equation}
	R(n) < R_L(n).
	\end{equation}
	Consequently, 
	$$ HIT < HIT_L.$$
\end{theorem}

\begin{theorem}[HIT of Event-Based Spreading Targeted Lockdown]\label{lockethm}
	Assume that an event-based spreading targeted lockdown was performed for $n \in [n_b, n_e]$. Then for any $n > n_e$,	
	\begin{equation}
	R_L(n) < R(n).
	\end{equation}
	Consequently, 
	$$ HIT > HIT_L.$$
\end{theorem}

Due to the generality of Theorems \ref{lockpthm} and \ref{lockethm}, we state the following corollary:

\begin{corollary}
	Any sequence of personal-trait spreading targeted lockdowns will result with $HIT < HIT_L$. 
	Similarly, any sequence of event-Based spreading targeted lockdowns will result with $HIT > HIT_L$.
\end{corollary}

The following claim will be useful in proving Theorems \ref{lockpthm} and \ref{lockethm}.
\begin{claim}\label{clm_fsd}
	Let $S_1, S_2$ be continuous random variables. Let $\rho_1(),\rho_2()$ be their pdfs, and $\P_1(),\P_2()$ their CDFs, respectively. Let $r()$ be an injective monotone function. If for any $s_1>s_2$: 
	$$\frac{\rho_2(s_1)}{\rho_1(s_1)}>\frac{\rho_2(s_2)}{\rho_1(s_2)}\:\:
	\text{ or }
	\:\:\frac{\rho_1(s_1)}{\rho_2(s_1)}<\frac{\rho_1(s_2)}{\rho_2(s_2)}$$
	then 
	$$\mathbb{E}[r(S_1)]\le\mathbb{E}[r(S_2)].$$
\end{claim}

\begin{proof}[Proof of Claim \ref{clm_fsd}]
	The proof is given in Appendix \ref{appendA2}.
\end{proof}

\subsection{Personal-Trait Lockdown: Proving Theorem \ref{lockpthm}}
Recall that $\rho(s,n)$ is the normalized susceptibility distribution (density) at step $n$ assuming a "natural" spread of the disease (i.e., no lockdown). 
Let $\rho_L(s,n)$ be the normalized susceptibility distribution (density) at step $n$ assuming that a lockdown was performed.

\begin{proof}[Proof of Theorem \ref{lockpthm}]
	Resulting from Lemma \ref{lem33} and by the assumption that $p_L = 0$ for $n_b \le n \le n_e$, at the end of the (personal-trait) lockdown, i.e. at $n_e$, it holds that
	\begin{equation}\label{rhoatend}
	\rho_{L}(s,n_e) = \rho_{L}(s,n_b) = \rho(s,n_b).
	\end{equation}
	However,
	\begin{equation}
	\rho(s,n_e)\approx\frac{\rho(s,n_{b})\cdot\exp\left(-\beta(n_b, n_e)\cdot(p\cdot s)\right)}{\int\rho(\sigma,n_{b})\cdot\exp\left(-\beta(n_b, n_e)\cdot(p\cdot\sigma)\right)d\sigma}.
	\end{equation}	
	Therefore,
	\begin{equation}
	\frac{	\rho_L(s,n_e)}{	\rho(s,n_e)} \approx 
	\exp\left(\beta(n_b, n_e)\cdot(p\cdot s)\right) \cdot c 
	\end{equation}	
	where $c = \int\rho(\sigma,n_{b})\cdot\exp\left(-\beta(n_b, n_e)\cdot(p\cdot\sigma)\right)d\sigma$.
	Therefore, for any $s_1 > s_2$:
	\begin{equation}\label{eqrsmallp}
	\frac{\rho_L(s_1,n_e)}{\rho(s_1,n_e)} > \frac{\rho_L(s_2,n_e)}{\rho(s_2,n_e)}.
	\end{equation}	
	Note that: $r(s) = (p \cdot s + q \cdot \lambda_{S})^2$ is an injective monotone function. Hence, by the definition of $R(n)$ (Eq. (\ref{R_expers})) and from Claim \ref{clm_fsd}:
	\begin{equation}\label{eqrsmall}
	R(n_e) < R_L(n_e).
	\end{equation}
	This is demonstrated in Figure \ref{figlock1}.
	
	\begin{figure}[h]
		\centering
		\includegraphics[width=\linewidth]{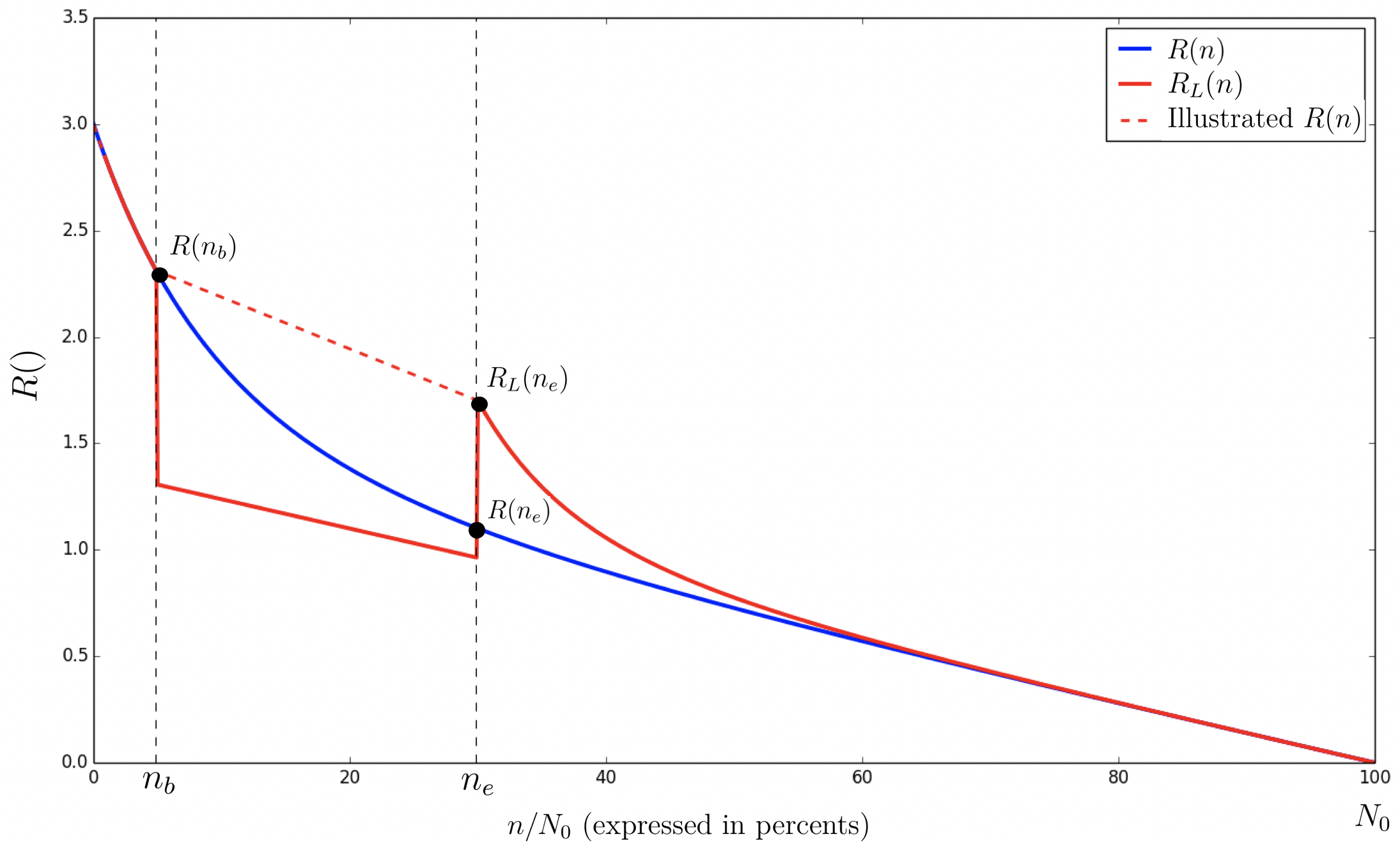}
		\caption{The expected value of $R_L(n)$ under a personal-trait spreading targeted lockdown (vs. $R(n)$ in a natural evolution) assuming $p=0.5$, $R_0=3$ and $k=0.1$. The lockdown begins at $n_b = 5\%$ and ends at $n_e = 30\%$.}
		\label{figlock1}
	\end{figure}
		
	More specifically, Since $\rho_{L}(s,n_e) = \rho(s,n_b)$ (Eq. (\ref{rhoatend})), and by Eq. (\ref{R_expers}), it holds that
	\begin{equation}\label{r_ratio}
	R_L(n_e)=\frac{N_0-n_e}{N_0-n_b}R(n_{b}).
	\end{equation}
	
	We move to measure the value of the effective reproduction number after the lockdown ends. Let $n > n_e$, and denote
	$$x=\frac{n-n_{e}}{N_0 - n_e}.$$
	Let
	$$n' = n_b + x \cdot (N_0 - n_b)$$
	(note that $n = n_e + x \cdot (N_0 - n_e)$). 
	$n$ and $n'$ are depicted in Figure \ref{figlock2}. 
	
	At the end of the lockdown $p$ returns to its original value and the disease spreads naturally, with its original parameters. At that point ($n_e$), according to Eq. (\ref{rhoatend}), the distribution of the population is the same as it was before the lockdown began, where only the size of the susceptible population was changed. 
	In other words, $R_L$ (red) behave to the right of $n_e$ exactly as $R$ (blue) behaves to the right of $n_b$ (see Figure \ref{figlock2}). 
	Therefore, 
	\begin{equation}\label{eqratior}
	\frac{R_L(n)}{R_L(n_{e})}=\frac{R(n')}{R(n_{b})}.
	\end{equation}
	This is demonstrated in Figure \ref{figlock2}. 
	\begin{figure}[h]
		\centering
		\includegraphics[width=\linewidth]{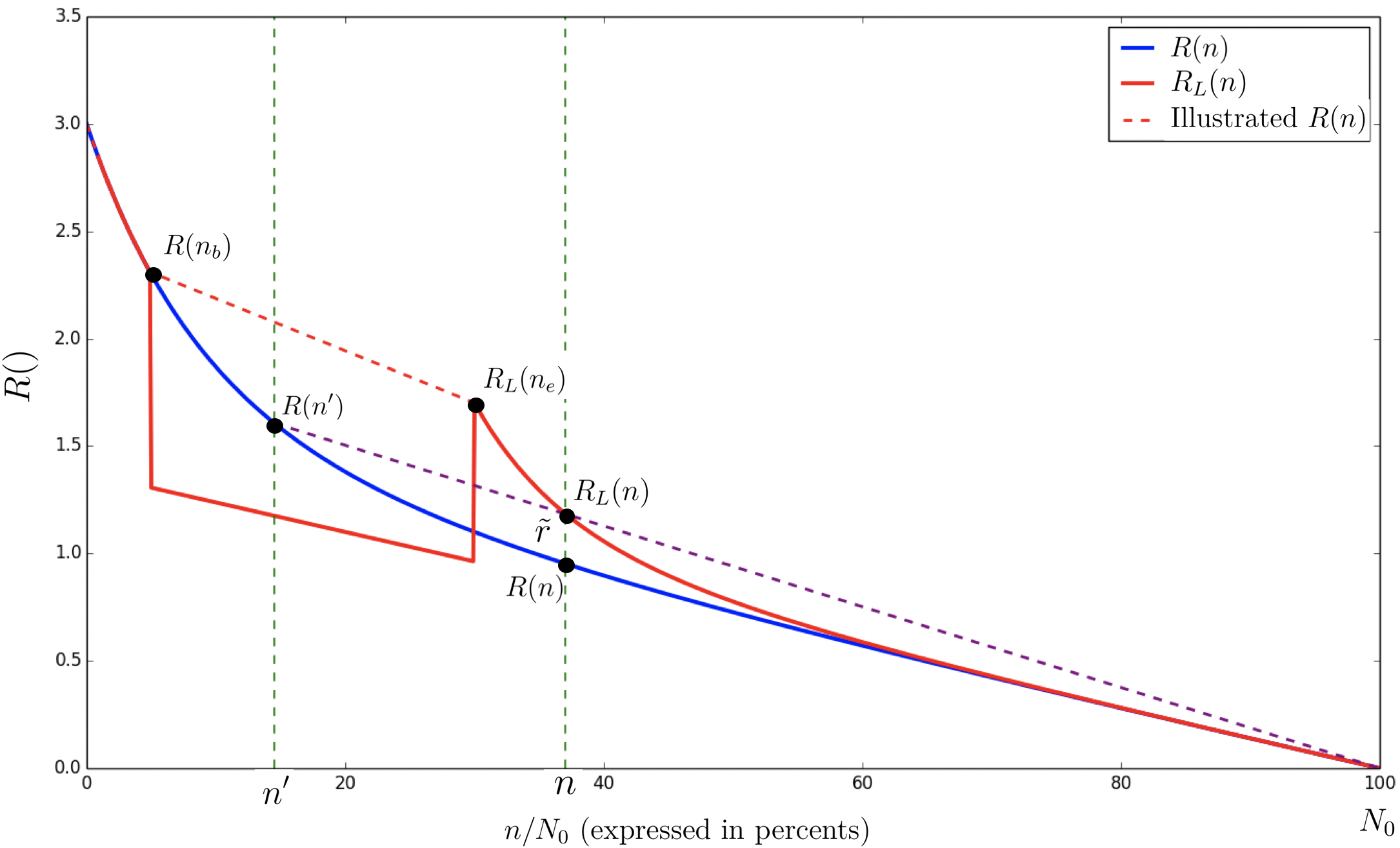}
		\caption{The expected value of $R_L(n)$ under a personal-trait spreading targeted lockdown (vs. $R(n)$ in a natural evolution) assuming $p=0.5$, $R_0=3$ and $k=0.1$. The lockdown begins at $n_b = 5\%$ and ends at $n_e = 30\%$. We use $x = 10\%$. The dashed purple line demonstrates a linear reduction in the value of $R(n')$. As proven, the purple and red curves intersect at $n$.}
		\label{figlock2}
	\end{figure}
	Combining Eq. (\ref{eqratior}) with Eq. (\ref{r_ratio}),
	\begin{equation}\label{eqeq}
	\frac{R_L(n)}{R(n')}=\frac{N_0-n_e}{N_0-n_b}.
	\end{equation}
	Let us look at 
	\begin{equation}\label{tilder}
	\tilde{r} \eqdef N(n) \cdot \int  \rho(\sigma, n') \cdot \left(p\cdot \sigma + q\cdot \lambda_{S} \right) \cdot \left(p\cdot \varphi(\sigma)+q\cdot \lambda_{I} \right) d\sigma.
	\end{equation}
	Hence,
	$$\frac{\tilde{r}}{R(n')} = \frac{N_0 - n}{N_0-n'}.$$
	Note that:
	$$
	\frac{N_0 - n}{N_0-n'} 
	= 
	\frac{N_0 - (n_e + x\cdot(N_0 - n_e))}{N_0 - (n_b + x\cdot(N_0 - n_b))}
	=
	$$
	$$ 
	=
	\frac{N_0 - n_e  -x \cdot N_0 + x \cdot n_e}{N_0 - n_b - x\cdot N_0  + x \cdot n_b}
	=
	\frac{(1 - x) \cdot (N_0 - n_e)}{(1 -x) \cdot (N_0 -n_b)}
	=
	\frac{N_0 - n_e}{N_0 -n_b}.
	$$
	Using Eq. (\ref{eqeq})
	$$\tilde{r}=R_L(n).$$
	In Eq. (\ref{tilder}) $\tilde{r}$ was calculated using $\rho(\sigma, n')$. Hence, as in Eq. (\ref{eqrsmallp}) and (\ref{eqrsmall}), 
	$$ R(n) < \tilde{r}, $$
	and we have that $R(n) < R_L(n)$. This holds for any $n > n_e$, and we conclude that $HIT < HIT_L$.
\end{proof}

Hence, while performing a personal-trait spreading targeted lockdown will reduce the value of the effective reproduction number during the lockdown, it might increase the number of individuals that contract with the disease, prior to reaching herd immunity.
In addition, note that even if "herd immunity" is reached during the lockdown (i.e., having $R_L(n) < 1$ for $n \in [n_b, n_e]$), it might not hold once the lockdown releases as the effective reproduction number might become greater than 1.

\subsection{Event-Based Lockdown: Proving Theorem \ref{lockethm}}
In the proof of Theorem \ref{lockethm} we use the following notation. Let
\begin{equation*}
\beta(n_1, n_2) = \sum_{i=n_1}^{n_2 - 1} \frac{1}{N(i) \cdot \mathbb{E}[S^i(H_i)]}.
\end{equation*}
I.e., $\beta(n_1, n_2) = \beta(n_2) - \beta(n_1)$ according to Eq. (\ref{betadef}). Let $\beta_L(n_1, n_2)$ be defined assuming that a lockdown was performed.
We prove the following claim:

\begin{claim}\label{clmlockrho}
	Under an event-base lockdown, for any $n \ge n_e$ and for any $s_1 > s_2$,
	\begin{equation}
	\frac{\rho(s_1,n)}{\rho_{L}(s_1,n)} > \frac{\rho(s_2,n)}{\rho_{L}(s_2,n)}.
	\end{equation}	
\end{claim}

\begin{proof}[Proof of Claim \ref{clmlockrho}]	
	When the lockdown begins, we have:
	\begin{equation}
	\rho(s,n_{b})=\rho_{L}(s,n_{b}).
	\end{equation}
	For any $n_b \le n \le n_e$,
	\begin{equation}
	\rho(s,n)\approx\frac{\rho(s,n_{b})\cdot\exp\left(-\beta(n_b, n)\cdot(p\cdot s)\right)}{\int\rho(\sigma,n_{b})\cdot\exp\left(-\beta(n_b, n)\cdot(p\cdot\sigma)\right)d\sigma}
	\end{equation}
	and
	\begin{equation}
	\rho_{L}(s,n)\approx\frac{\rho_{L}(s,n_{b})\cdot\exp\left(-\beta_{L}(n_b, n)\cdot(p\cdot s)\right)}{\int\rho_{L}(\sigma,n_{b})\cdot\exp\left(-\beta_{L}(n_b, n)\cdot(p\cdot\sigma)\right)d\sigma}.
	\end{equation}
	The ratio between the density functions is: 
	$$\frac{\rho(s,n)}{\rho_{L}(s,n)}=
	\frac{\rho(s,n_b)}{\rho_{L}(s,n_b)}\cdot\frac{\exp\left(-\beta(n_b, n)\cdot(p\cdot s)\right)}{\exp\left(-\beta_{L}(n_b, n)\cdot(p\cdot s)\right)}\cdot\frac{\int\rho_{L}(\sigma,n)\cdot\exp\left(-\beta_{L}(n_b, n)\cdot(p\cdot\sigma)\right)d\sigma}{\int\rho(\sigma,n)\cdot\exp\left(-\beta(n_b, n)\cdot(p\cdot\sigma)\right)d\sigma}.$$

	Hence:
	\begin{equation}\label{eq1ofclmlock}
	\frac{\rho(s,n)}{\rho_{L}(s,n)} = \exp(p \cdot s \cdot (\beta_{L}(n_b, n) - \beta(n_b, n))\cdot c
	\end{equation}
	where $c = \frac{\int\rho_{L}(\sigma,n_{e})\cdot\exp\left(-\beta_{L}(n_b, n_{e})\cdot(p\cdot\sigma)\right)d\sigma}{\int\rho(\sigma,n_{e})\cdot\exp\left(-\beta(n_b, n_{e})\cdot(p\cdot\sigma)\right)d\sigma}$ .
	We establish the following claim.
	\begin{claim}\label{helplem1}
		For any $n_b < n \le n_e$,
		$$\beta_{L}(n_b, n)>\beta(n_b, n).$$ 
	\end{claim}	
	\begin{proof}[Proof of Claim \ref{helplem1}]
		
		We will prove the claim using induction on $n$. 
		\begin{itemize}
			
			\item (Base case). Since $q_L = 0$ ( $< q$) we have that: $S_L^{n_b}(a) < S^{n_b}(a)$ for any $a$. Hence,  $\beta_L(n_b, n_b+1) > \beta(n_b, n_b+1)$.
			\item (Inductive step). Let $n\in(n_b, n_e)$. Assume that 
			$$\beta_{L}(n_b, {n}) > \beta(n_b, {n}).$$
			By Eq. (\ref{eq1ofclmlock}), for any $s_1 > s_2$,
			$$ \frac{\rho(s_1,n)}{\rho_{L}(s_1,n)} > \frac{\rho(s_2,n)}{\rho_{L}(s_2,n)},$$
			and using Claim \ref{clm_fsd}, 
			$$\mathbb{E}[S_L^n(a)] < \mathbb{E}[S^n(a)]$$ 
			Using the definition of $\beta$ and by our assumption,
			$$\beta_L(n_b, n+1) > \beta(n_b, n+1).$$
		\end{itemize}
		we conclude that for any $n_b < n \le n_e$, $\beta_{L}(n_b, n)<\beta(n_b, n)$.
	\end{proof}

	Next, we apply Claim \ref{helplem1} for $n_e$ and have that $\beta_{L}(n_b, n_e) - \beta(n_b, n_e) > 0$. Therefore, by Eq. (\ref{eq1ofclmlock}) for any $s_1 > s_2$
	\begin{equation}\label{forhelplem2}
	\frac{\rho(s_1,n_e)}{\rho_{L}(s_1,n_e)} > \frac{\rho(s_2,n_e)}{\rho_{L}(s_2,n_e)},
	\end{equation}
	and according to Claim \ref{clm_fsd}, $R_L(n_e) < R(n_e)$, where $R$ and $R_L$ are calculated using the original values of $p$ and $q$. This is demonstrated in Figure \ref{figlock3}.

	\begin{figure}[h]
		\centering
		\includegraphics[width=\linewidth]{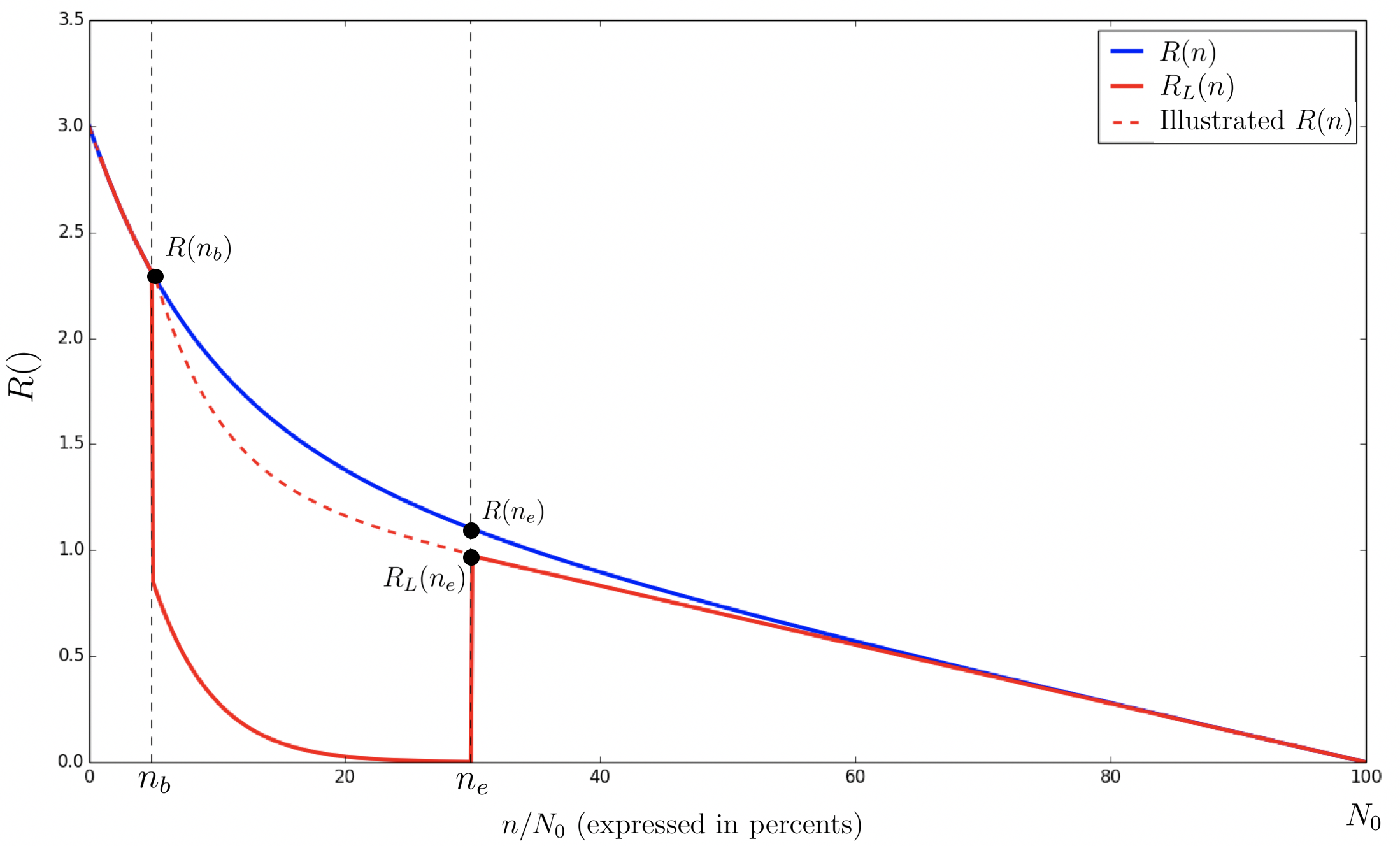}
		\caption{The expected $R_L(n)$ under an event-based spreading targeted lockdown (vs. $R(n)$ in a natural evolution) assuming $p=0.5$, $R_0=3$ and $k=0.1$. The lockdown begins at $n_b = 5\%$ and ends at $n_e = 30\%$.}
		\label{figlock3}
	\end{figure}
	
	In a similar develop as in Eq. (\ref{eq1ofclmlock}), we have that for any $n > n_e$,
	\begin{equation}\label{helpeq1}
	\frac{\rho(s,n)}{\rho_{L}(s,n)} =  \exp(p \cdot s \cdot ( (\beta_{L}(n_b, n_{e}) +  \beta_{L}(n_e, n)) - (\beta(n_b, n_e) + \beta(n_e, n)) )\cdot c \cdot c'.
	\end{equation}	
	where  $c' = \frac{\int\rho_{L}(\sigma,n)\cdot\exp\left(-\beta_{L}(n_e, n)\cdot(p\cdot\sigma)\right)d\sigma}{\int\rho(\sigma,n)\cdot\exp\left(-\beta(n_e, n)\cdot(p\cdot\sigma)\right)d\sigma}$ .
	We establish the following claim, similar to Claim \ref{helplem1}:
	
	\begin{claim}\label{helplem2}
		For any $n_e < n$,
		$$\beta_{L}(n_e, n)>\beta(n_e, n).$$ 
	\end{claim}	
	\begin{proof}[Proof of Claim \ref{helplem2}]
		The proof follows the same idea of the proof of Claim \ref{helplem1}.The full proof is given in Appendix \ref{appendA2}.
	\end{proof}
	We conclude the proof of the claim using Claim \ref{helplem2} and Eq. (\ref{helpeq1}). It holds that $\forall s_1 > s_2$, $\forall n \ge n_e$,
	$$\frac{\rho(s_1,n)}{\rho_{L}(s_1,n)} > \frac{\rho(s_2,n)}{\rho_{L}(s_2,n)}.$$ 
\end{proof}
\begin{proof}[Proof of Theorem \ref{lockethm}]
	Combining Claim \ref{clmlockrho} with Claim \ref{clm_fsd} we have that for any $n \ge n_e$:
	\begin{equation}
	R_L(n) < R(n).
	\end{equation}
	Consequently, $HIT_L < HIT$, and Theorem \ref{lockethm} follows.
\end{proof}.

Note that the proof of Theorem \ref{lockethm} was not based on the assumption that $q_L = 0$, but on the fact that $q_L < q$ (see the proof of Claim \ref{helplem1}). 
Hence, Theorem \ref{lockethm} can be generalized to hold for any $0 \le q_L < q$, namely for \textit{Partial Lockdown}. 

\begin{corollary}[Generalization of Theorem \ref{lockethm}]\label{lockethm2}
	Any event-based spreading targeted partial lockdown, namely where $p$ remains at the same level and $q_L < q$, 	results in $ HIT_L < HIT$.
\end{corollary}

\begin{remark}[Generalization of Theorem \ref{lockpthm}]\label{lockpthm2}
	We conjecture that the generalization of Theorem \ref{lockpthm} for personal-trait spreading targeted partial lockdown holds as well.
\end{remark}

\section{Conclusions  and Discussion }
In this work we studied the effects of infectiousness heterogeneity (overdispersion) and lockdowns on herd immunity. Recent literature suggests that COVID-19 is characterized by such heterogeneity which affects dramatically herd immunity. We proposed that infectiousness (and susceptibility) should be classified into two inherently different types which we called \textit{personal-trait} and \textit{event-based} spreading.

We followed up \cite{oz2020heterogeneity} and proposed a model that accounts for both of these types. 
Under this new model we showed that herd immunity and the HIT strongly depend on the mix between the two types of spreading. 
We analyzed the decay of the effective reproduction number, $R()$, and showed that the contribution of the personal-trait spreading drops sharply at early stages of the disease while the contribution of the event-based spreading drops much more slowly.
That is - the \textit{super-spreaders} "leave the game" at early stages, while the \textit{super-spreading} events remain active.

We demonstrated the results on a (Gamma) distribution which previously attributed to the infectiousness of COVID-2 \cite{smith2005} and COVID-19 \cite{endo2020estimating,oz2020heterogeneity}.
We showed that in order to predict the mix between the spreading types, only a small fraction of the population is needed to get infected. Such prediction can help in calculating the HIT in an early stage of the disease. This estimation problem for COVID-19 is left open for forthcoming research. 

We addressed operational aspects of disease blocking and analyzed the effect of lockdowns on the HIT. 
We showed that different lockdown strategies, targeting different spreading types, result in opposite effect on the HIT. 
In particular, 
a lock-down which targets personal-trait spreading would act adversely and \textit{reduces}   herd immunity.  
This seems to fit a lockdown that focuses on daily/professional activities. 
In contrast, a lock-down which targets event-based spreading will \textit{increase} herd immunity. 
This may fit a lockdown that focuses on sports/social events. 

Of course, a lockdown may have other objectives such as achieving temporary slow down of the disease spreading (to allow handling the patients masses) which may justify the lockdown strategy. Yet -- the effect on herd immunity requires consideration, especially due to the opposite effects of various lockdowns. 

\bibliographystyle{abbrvnat}
\bibliography{CovidJTHL}

\appendix

\section{Proofs}\label{appendA2}

\begin{proof}[Proof of Claim \ref{lem32} (continued)]
	The proof follows the proof of Claim I provided in \cite{oz2020heterogeneity}. Taking natural $\log$ of Eq. (\ref{eqclmpra}),
	$$\log(\Pr[a\text{ is healthy at round \ensuremath{n}]})-\log(\Pr[a\text{ is healthy at round \ensuremath{n-1}}])=$$
	\begin{equation}
	=\log\text{\ensuremath{\left(1-\frac{S^{n-1}(a)}{N(n-1) \cdot \mathbb{E}_{b \sim H_{n-1}}[S^{n-1}(b)]}\right)}}.
	\end{equation}
	It holds that:
	$$ \log \left(1-\frac{S^{n-1}(a)}{N(n-1) \cdot \mathbb{E}_{b \sim H_{n-1}}[S^{n-1}(b)]}\right) =$$
	$$=-\frac{S^{n-1}(a)}{N(n-1) \cdot \mathbb{E}_{b \sim H_{n-1}}[S^{n-1}(b)]}-O\left(\left(\frac{S^{n-1}(a)}{N(n-1) \cdot \mathbb{E}_{b \sim H_{n-1}}[S^{n-1}(b)]}\right)^{2}\right).$$
	We attempt to find the herd-immunity threshold, and hence can bound number of steps, $n$, by $(1 - 1/R_0) \cdot N_0$. Hence, 
	$$ \log \left(1-\frac{S^{n-1}(a)}{N(n-1) \cdot \mathbb{E}_{b \sim H_{n-1}}[S^{n-1}(b)]}\right) =$$
	$$= 
	-\frac{S^{n-1}(a)}{N(n-1) \cdot \mathbb{E}_{b \sim H_{n-1}}[S^{n-1}(b)]}
	-
	O\left(\left(\frac{\max_b S^{n-1}(b)}{N_0 \cdot \mathbb{E}_{b}[S^{n-1}(b)]}\right)^{2}\right)
	.$$
	Counting over the steps $1,\dots,n$ we have:
	$$\log(\Pr[a\text{ is healthy at round \ensuremath{n}]}) = $$
	$$ -\sum_{i=1}^{n-1} \frac{S^{i}(a)}{N(i) \cdot \mathbb{E}_{b \sim H_{i}}[S^n(b)]} 
	- O\left(\left(\frac{\max_b S^n(b)}{\mathbb{E}_{b}[S^n(b)]}\right)^{2} \cdot \frac{n}{N_0^2} \right)
	.$$
	And as long as $\frac{\max_b S^n(b)}{\mathbb{E}_{b}[S^n(b)]} \ll \sqrt{N_0}$ we have that:
	
	\begin{equation}
	\log(\Pr[a\text{ is healthy at round \ensuremath{n}]}) \approx 
	- \sum_{i=1}^{n-1} \frac{S^{i}(a)}{N(i) \cdot \mathbb{E}_{b \sim H_{i}}[S^n(b)]}.
	\end{equation}
	Note that for any $i$,
	$$S^{i}(a) = p \cdot S_p(a) + q \cdot S_e^i(a).$$
	Since our calculation is done by taking an expectation over all possible scenarios of infections, we have:
	\begin{equation}
	\log(\Pr[a\text{ is healthy at round \ensuremath{n}]}) \approx -\beta(n) \cdot \left( p \cdot S_p(a) + q \cdot \lambda_{S}  \right)
	\end{equation}
	Hence,
	\begin{equation}
	\Pr[a\text{ is healthy at round \ensuremath{n}]} \approx \exp\left(-\beta(n)\cdot(p\cdot S_{p}(a)+q\cdot\lambda_{S})\right)
	\end{equation}
	and the proof is complete.
\end{proof}

\begin{proof}[Proof of Claim \ref{clm_fsd}]
	Let $s'>0$. By our assumption, for any $s<s'$:
	$$\rho_2(s')\rho_1(s)>\rho_2(s)\rho_1(s').$$
	Hence, integrating $s$ over $(0,s')$ we have:
	$$\int_{0}^{s'}\rho_2(s')\rho_1(s)-\rho_2(s)\rho_1(s')ds>0.$$
	Therefore:
	\begin{equation}\label{lemeq1}
	\rho_2(s')\P_1(s')=\int_{0}^{s'}\rho_2(s')\rho_1(s)ds>\int_{0}^{s'}\rho_2(s)\rho_1(s')ds=\P_2(s')\rho_1(s').
	\end{equation}
	By the same ideas we have that:
	\begin{equation}\label{lemeq2}
	(1-\P_2(s'))\rho_1(s')=\int_{s'}^{\infty}\rho_2(s)\rho_1(s')ds>\int_{s'}^{\infty}\rho_2(s')\rho_1(s)ds=\rho_2(s')(1-\P_1(s')).
	\end{equation}
	Using Eq. (\ref{lemeq1}) and Eq. (\ref{lemeq2}) we have:
	$$\frac{1-\P_2(s')}{1-\P_1(s')}>\frac{\rho_2(s')}{\rho_1(s')}>\frac{\P_2(s')}{\P_1(s')}$$
	and hence:
	\begin{equation}\label{fsdeq}
	\forall s,\ \P_2(s)<\P_1(s).
	\end{equation}
	Since $r$ is an injective monotone function, it holds that: 
	\begin{equation}\label{eqpr1}
	\Pr[S_1>s']=\Pr[r(S_1)>r(s')].
	\end{equation}
	In the same way:
	\begin{equation}\label{eqpr2}
	\Pr[S_2>s']=\Pr[r(S_2)>r(s')].
	\end{equation}
	Using Eq. (\ref{fsdeq}) combined with Eq. (\ref{eqpr1}) and \ref{eqpr2}):
	\begin{equation}\label{eqpr3}
	\Pr[r(S_1)>r(s')]<[r(S_2)>r(s')].
	\end{equation}
	For any non-negative random variable $X$ it holds that:
	\begin{equation}\label{eqexpect}
	\mathbb{E}[X]=\int_{0}^{\infty}1-F_{X}(s)ds.
	\end{equation}
	Therefore, by Eq. (\ref{eqpr3}) combined with \ref{eqexpect} we know that:
	$$\mathbb{E}[r(S_1)]\le\mathbb{E}[r(S_2)].$$
	And the proof is complete.
\end{proof}

\begin{proof}[Proof of Claim \ref{helplem2}]
	We will prove the claim using induction on $n$.
	\begin{itemize}
		
		\item (Base case). Using Eq. (\ref{forhelplem2}), we have that:
		$$\mathbb{E}[S_L^{n_e}(a)] < \mathbb{E}[S^{n_e}(a)].$$ 
		Hence, 	
		$$\beta_L(n_e, n_e+1) > \beta_L(n_e, n_e+1).$$	
		\item (Inductive step). Let $n_e < n$. Assume that 
		$$\beta_{L}(n_e, n) > \beta(n_e, n).$$
		By Claim \ref{helplem1} and Eq. (\ref{helpeq1}), for any $s_1 > s_2$,
		$$ \frac{\rho(s_1,n)}{\rho_{L}(s_1,n)} > \frac{\rho(s_2,n)}{\rho_{L}(s_2,n)},$$
		and using Claim \ref{clm_fsd}, 
		$$\mathbb{E}[S_L^n(a)] < \mathbb{E}[S^n(a)]$$ 
		Using the definition of $\beta$ and by our assumption,
		$$\beta_L(n_e, n+1) > \beta(n_e, n+1).$$
	\end{itemize}
	we conclude that for any $n_e < n$, $\beta_{L}(n_e, n)<\beta(n_e, n)$.
\end{proof}

\end{document}